\documentclass[letter,twocolumn]{jpsj2}

\usepackage{graphicx}

\def\Tc{T_\mathrm{c}}

\def\Sig{\mathit{\Sigma}}

\def\Del{\mathit{\Delta}}

\def\Re{\mathrm{Re}}

\def\G{\mathcal{G}}
\def\F{\mathcal{F}}
\def\lamx{\lambda_\mathrm{max}}

\title{%
Possibility of $f$-wave spin-triplet superconductivity 
in the CoO superconductor: \\
a case study on a 2D triangular lattice in the repulsive Hubbard model
}

\author{%
Hiroaki \textsc{Ikeda}$^1$
\thanks{E-mail:hiroaki@scphys.kyoto-u.ac.jp},
Yunori \textsc{Nisikawa}$^2$ 
and Kosaku \textsc{Yamada}$^1$
}

\inst{
$^1$Department of Physics, Kyoto University, Sakyo-ku, Kyoto 606-8502 \\
$^2$Synchrotron Radiation Research Center,
Japan Atomic Energy Research Institute, Mikazuki, Sayo, Hyogo 679-5148
}

\recdate{\today}

\abst{%
Stimulated by the recent finding of Na$_{0.35}$CoO$_2$.1.3H$_2$O 
superconductor, we investigate superconducting instabilities 
on a 2D triangular lattice in the repulsive Hubbard model.
Using the third-order perturbation expansion with respect to the on-site 
repulsion $U$, we evaluate the linearized Dyson-Gor'kov equation.
We find that an $f$-wave spin-triplet pairing is the most stable
in a wide range of the next nearest neighbor hopping integral $t'$ and 
an electron number density $n$.
The introduction of $t'$ is crucial to adjust the van Hove singularities
to the neighborhood of the Fermi surface crossing around K point.
In this case, the bare spin susceptibility shows the broad peak 
around $\Gamma$ point.
These conditions stabilize the $f$-wave pairing.
Although the $f$-wave pairing is also given by the fluctuation-exchange 
approximation, the transition temperature is too low to be observed.
This is because the depairing effect by the spin fluctuation is over-estimated.
Thus, the third-order vertex corrections are important 
for the spin-triplet superconductivity,
like the case in Sr$_2$RuO$_4$.
}

\kword{%
Na$_x$CoO$_2$.$y$H$_2$O, $f$-wave spin-triplet superconductivity, 
third-order perturbation theory, fluctuation exchange
}

\begin{document}
\sloppy
\maketitle

Recently, superconductivity in Na$_{0.35}$CoO$_2$.1.3H$_2$O with 
the transition temperature $\Tc\simeq 5$K has been discovered~\cite{rf:Takada}
and intensively investigated.
This compound has a hexagonal layered structure, and the dominant 
conductive plane is formed by the network of the CoO$_2$ octahedra.
It is considered as the 2D layered triangular lattice.
The deintercalation of Na$^+$ ions corresponds to 0.65/Co hole doping 
for Co$^{3+}(3d)^6$ with fulfilled $t_{2g}$ orbits
in the CoO$_2$ layer.
The uniform susceptibility shows the weak-ferromagnetic like behavior, and
indicates relevance of electron correlations.
Up to now, although the NMR/NQR measurements has been performed 
in three groups,~\cite{rf:Waki,rf:Kobayashi,rf:Fujimoto}
they have contradicted each other about the pairing symmetry 
of the superconductivity.
However, absence of the coherence peak just below $\Tc$ observed 
by Fujimoto {\it et al.}~\cite{rf:Fujimoto} has suggested possibility of 
unconventional superconductivity with line-node gap.
Quite recently, this experimental fact has been confirmed
by Ishida {\it et al.}~\cite{rf:Ishida}
Thus, we are naturally interested to know what kind of superconductivity
can be caused by the correlated electrons on the 2D triangular lattice.

About such superconducting instabilities on the triangular lattice, 
several theoretical studies have been already reported.
In RVB superconductivity in the framework of the {\it t-J} 
model,~\cite{rf:Baskaran,rf:Kumar,rf:Wang,rf:Ogata}
and in a study using a one-loop renormalization group,~\cite{rf:Honer}
a $d$+i$d$-wave spin-singlet pairing has been discussed actively.
Possibility of a $p$+i$p$-wave spin-triplet pairing has been proposed
phenomenologically.~\cite{rf:Tanaka}
Within the paramagnon exchange, a $d$-wave singlet and an odd-frequency 
$s$-wave triplet pairings has been discussed.~\cite{rf:Vojta}
In a fluctuation-exchange (FLEX) approximation, possibility of an $f$-wave 
triplet pairing at low carrier limit has been also suggested.~\cite{rf:Kuroki}

So far, we have investigated unconventional superconductivity 
in repulsive Hubbard models for many superconductors including
the high-$\Tc$ cuprates and Sr$_2$RuO$_4$
on the basis of the third-order perturbation theory.~\cite{rf:Review}
All of them are in agreement with experimental facts, 
especially, concerning the pairing symmetry.
Thus, here, in the same framework, we investigate superconducting 
instabilities in a simple one-band Hubbard model on the 2D triangular lattice.
In this one-band model, 0.65/Co hole doping in Na$_{0.35}$CoO$_2$.1.3H$_2$O
corresponds to 1.35 electron filling, or 0.65 electron filling 
by the particle-hole transformation if the band structure is upside-down.
Thus, the results for the electron number density $n=$1.35 or 0.65 
in this model will be helpful for the study of the CoO$_2$ superconductor.

Now, let us consider the repulsive Hubbard model on the 2D triangular lattice,
\begin{equation}
H=\sum_k \xi_k c_{k\sigma}^\dagger c_{k\sigma}
+ U\sum_i n_{i\uparrow}n_{i\downarrow}.
\end{equation}
The dispersion relation is given by
\begin{equation}
\begin{split}
\xi_k =& -2t\Bigr(
2\cos\frac{\sqrt{3}k_x}{2}\cos\frac{k_y}{2}+\cos k_y
\Bigr) \\
& -2t'\Bigr(
\cos\sqrt{3}k_x+2\cos\frac{\sqrt{3}k_x}{2}\cos\frac{3k_y}{2}
\Bigr) -\mu,
\end{split}
\end{equation}
where $t$ and $t'$ are the nearest neighbor and the next nearest neighbor
hopping integrals, respectively.
The chemical potential $\mu$ is determined by fixing $n$.
In this system, we apply the third-order perturbation expansion 
with respect to $U$, and evaluate the linearized Dyson-Gor'kov 
(gap) equation.~\cite{rf:calc}
The normal self-energy is given by
\begin{subequations}
\begin{align}
\label{eq:sig2}
\Sig(k)&= \sum_{k'} U^2 \chi_0(k-k')\G_0(k') \\
& +U^3 \left( \chi_0^2(k-k')+\phi_0^2(k+k') \right) \G_0(k'),
\end{align}
\end{subequations}
where $\chi_0(q),\phi_0(q)=-\sum_p \G_0(p)\G_0(q\pm p)$.
Since the first-order normal self-energy is constant and can be included 
by the chemical potential $\mu$, here we omit it.
The gap equation is given by
\begin{subequations}
\begin{align}
&\lambda\Del(k)=-\sum_{k'}U^2\chi_0(k-k')\F(-k') \\
&-\sum_{k'}U^3{\chi_0(k-k')}^2 \left(\F(k')+\F(-k')\right) \\
&-\sum_{k'q}U^3 2\Re\left[\chi(q)\G_0(k+q)\G_0(k'+q)\right]\F(k') \\
&+\sum_{k'q}U^3 2\Re\left[\phi(q)\G_0(k+q)\G_0(k'+q)\right]\F(-k'),
\end{align}
\label{eq:del}
\end{subequations}
for both the spin-triplet and singlet pairs,~\cite{rf:gapeq}
where $\F(k)=|\G(k)|^2\Del(k)$, and $\lambda$ denotes an eigen value 
for each eigen vector $\Del(k)$.
The temperature with $\lambda=1$ corresponds to $\Tc$.
By estimating the largest positive eigen value $\lamx$ in the gap equation,
we determine which kind of symmetry is stable among irreducible representations
(IR) of the space group $D_{6h}$: 
\begin{tabular}{cc}
\hline 
IR (symmetry) & One of Basis funcitons \\
\hline
$B_1$ ($f$-wave) & $\sin(\frac{k_y}{2})
(\cos(\frac{\sqrt{3}k_x}{2})-\cos(\frac{k_y}{2}))$ \\
$B_2$ ($f$-wave) & $\sin(\frac{\sqrt{3}k_x}{2})
(\cos(\frac{\sqrt{3}k_x}{2})-\cos(\frac{3k_y}{2}))$ \\
$E_1$ ($p$-wave) & $\sin(\frac{\sqrt{3}k_x}{2})\cos(\frac{k_y}{2})$ \\
$E_2$ ($d$-wave) & $\sin(\frac{\sqrt{3}k_x}{2})\sin(\frac{k_y}{2})$ \\
\hline
\end{tabular}

{\it Less than half} ---
\begin{figure}[t]
\begin{center}
\includegraphics[width=6cm]{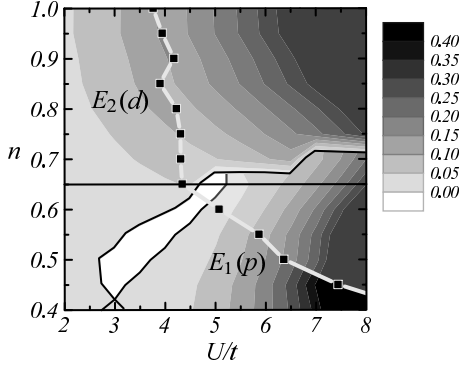}
\end{center}
\caption{The contour plot of eigen values for $0.4<n<1.0$ and $2<U/t<8$
at $t'=0$ and $T/t=0.01$. 
The gray thick line with points denotes $1/\chi_0^\mathrm{max}$
as the reference value of $U$.
The $d$-wave for $0.7\lesssim n \lesssim 1.0$ and the $p$-wave 
for $0.4\lesssim n\lesssim 0.6$ are stable.}
\label{fig:U-n}
\end{figure}
First of all, let us consider the case of $n<1.0$ less than half filling.
In this case, the Fermi surface is round, and the density of states is low.
Since the $t'$-dependence also is weak, we set $t'=0$.
In Fig.\ref{fig:U-n}, we illustrate the contour plot of $\lamx$ at $T/t=0.01$.
The white and black regions correspond to $\lamx<0.0$ and $\lamx>0.4$,
respectively.
The gray thick line with points represents $1/\chi_0^\mathrm{max}$ 
as a reference value of $U$, where $\chi_0^\mathrm{max}$ is 
the maximum value of the bare susceptibility $\chi_0(q,0)$.
Thus, actually, $U/t$ may be too large to realize the faded region
on the right-hand side of this line.
Near the half filling, the $d$-wave singlet pairing ($E_2$) is stable.
Far from the half filling, it becomes unstable, and possitive eigen values
are obtained only for small $U$.
Rather, the $p$-wave triplet pairing ($E_1$) is stable for large $U$.
Thus, in this region, the convergence of solutions with respect to $U$ 
is not good.
The second-order (Eq.(\ref{eq:del}a)) and 
the third-order (Eq.(\ref{eq:del}b-d)) scattering processes 
prefer different pairing states; the former the singlet, the later the triplet.
However, if assuming that $U \sim 1/\chi_0^\mathrm{max}$, 
the $d$-wave for $0.7\lesssim n \lesssim 1.0$ and the $p$-wave 
for $0.4\lesssim n\lesssim 0.6$ are stable, 
and $0.6\lesssim n\lesssim 0.7$ is on the border line.
The electron filling $n=0.65$ corresponding to Na$_{0.35}$CoO$_2$.1.3H$_2$O 
is on this border line.
Thus, we can not expect the stable superconductivity at $\Tc \sim 5$K 
within this analysis.

{\it More than half} ---
\begin{figure}[t]
\begin{center}
\includegraphics[width=6cm]{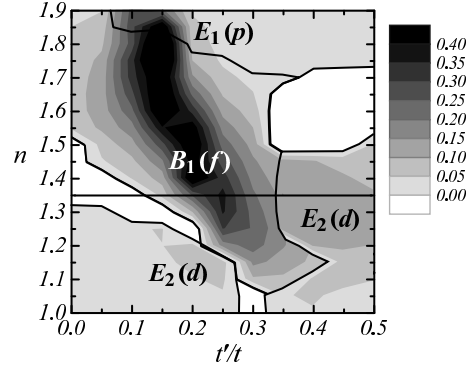}
\end{center}
\caption{The phase diagram of $n$ and $t'$ at $U/t=1.8$ and $T/t=0.01$. 
The eigen values for the black region is above 0.4.
The $f$-wave triplet pairing ($B_1$) is the most stable in the wide region.}
\label{fig:n-t}
\end{figure}
Next, we consider the case of $n>1.0$ more than half filling.
This situation is sensitive to introduction of $t'$.
Fig.\ref{fig:n-t} shows the contour plot of $\lamx$ 
at $U/t=1.8$ and $T/t=0.01$.
The solid line with $\lamx=0.0$ distinguishes four regions.
Two $d$-wave singlet ($E_2$) regions locate near the half filling $n=1.0$.
These regions are continuously connected with the region of $n$ 
less than half filling.
In this case, $\lamx$ is less than 0.15 at most.
On the other hand, far from the half filling, the $f$-wave triplet pairing 
($B_1$) has relatively large $\lamx$, and covers a wide region
of the phase diagram.
Successively, the region of $p$-wave triplet ($E_1$) spreads
only near the low carrier $n\sim 1.9$.
This indicates that the $f$-wave triplet superconductivity 
can be realized rather than the $d$-wave singlet one
at the carrier number far from the half filling on the 2D triangular lattice.
This is also the case on the density $n=1.35$ corresponding to 
Na$_{0.35}$CoO$_2$.$1.3$H$_2$O.
The pairing state with $B_1$ symmetry is the most stable with large $\lamx$ 
in $0.12<t'/t<0.32$.
Although the pairing state with $E_2$ symmetry possesses the positive 
eigen value in $t'/t>0.32$, we can not expect so high $\Tc$.
Such a difference originates from the location of the van Hove singularities
near the Fermi surface.
Let us discuss the situation in details below.

\begin{figure}[t]
\begin{center}
\includegraphics[width=8cm]{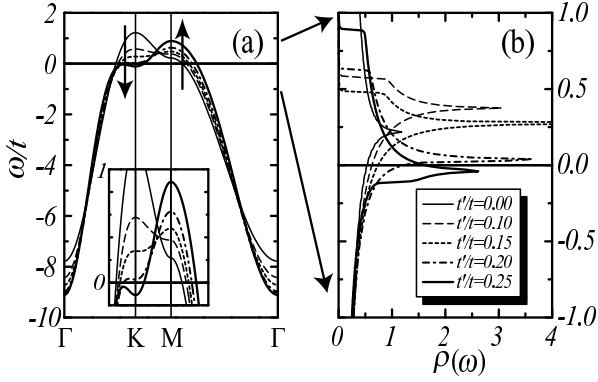}
\end{center}
\caption{(a) The dispersion relation $\xi_k$ along the symmetry line 
and (b) The density of states $\rho(\omega)$ at $n=1.35$ 
for $t'/t=0.00$, 0.10, 0.15, 0.20 and 0.25.
The symmetry line is indicated in Fig.\ref{fig:fermi}(c).
The inset in (a) is the enlarged figure of the neighborhood 
of the Fermi level.}
\label{fig:dos}
\end{figure}

\begin{figure}[t]
\begin{center}
\includegraphics[width=6.5cm]{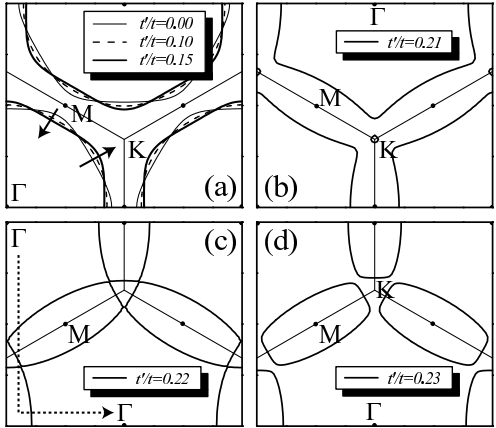}
\end{center}
\caption{The Fermi surfaces at $n=1.35$ for (a) $t'/t=0.00$, 0.10, 0.15,
(b) $t'/t=0.21$, (c) $t'/t=0.22$ and (d) $t'/t=0.23$.}
\label{fig:fermi}
\end{figure}

\begin{figure}[t]
\begin{center}
\includegraphics[width=8cm]{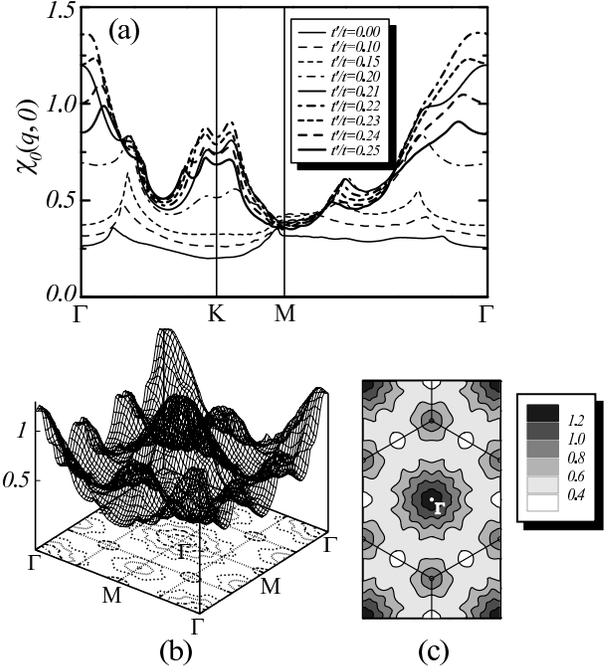}
\end{center}
\caption{(a) The bare spin susceptibility $\chi_0(q,0)$ along 
the symmetry line at $n=1.35$ and $T/t=0.01$ for $t'/t=0.00$, 
0.10, 0.15, 0.20, 0.21, 0.22, 0.23, 0.24 and 0.25.
(b) $\chi_0(q,0)$ at $t'/t=0.23$ and (c) the contour plot.
We can see peak or hump structures around $\Gamma$ and K points.}
\label{fig:chi}
\end{figure}

In Fig.\ref{fig:dos}, we illustrate the dispersion relation $\xi_k$ and 
the density of states $\rho(\omega)$ at $n=1.35$.
At $t'=0$, the van Hove points exist at M point, and 
K point is the maximum of $\xi_k$.
This situation is sensitive to the introduction of $t'$.
With increasing $t'$, the energy at K point decreases, and 
that at M point increases.
Correspondingly, the van Hove points move around K point, and 
the singularities cross the Fermi level at $t'/t \sim 0.21$.
We can also verify it by $t'$-dependence of the peak position 
in $\rho(\omega)$.
Since the Fermi level is located near the van Hove singularities,
the shape of the Fermi surface is also sensitive to $t'$.
With increasing $t'$, the shape changes from Fig.\ref{fig:fermi}(a) 
to Fig.\ref{fig:fermi}(d).
In Fig.\ref{fig:chi}, we display the bare spin susceptibility $\chi_0(q,0)$ 
along the symmetry line and the typical structure at $t'/t=0.23$
corresponding to Fig.\ref{fig:fermi}(d).
The introduction of $t'$ leads to the remarkable wave-vector dependence:
the incommensurate peak or the broad peak structure near $\Gamma$ point,
and the hump structure near K point.
The enhancement of $\chi_0(0,0)$ at $\Gamma$ point corresponds to 
the increase of the density of states by the van Hove singularity.
On the other hand, the hump structure near K point originates from 
the peculiarity of the crystal structure.
The wave-vector at K point in $\chi_0(q,0)$ links not only $\Gamma$ point 
and K point in the energy dispersion, but also the neighboring K points.
Thus, the hump structure also reflects the effect of the van Hove points 
around K point in $\xi_k$.
The peak structure at $\Gamma$ point in $\chi_0(q,0)$ implies 
the ferromagnetic instability near $t'/t\sim 0.21$.
Thus, the triplet superconductivity with $B_1$ symmetry may be located
in the neighborhood of the ferromagnetic phase,
when we take the magnetic instability into account.

\begin{figure}[t]
\begin{center}
\includegraphics[width=8cm]{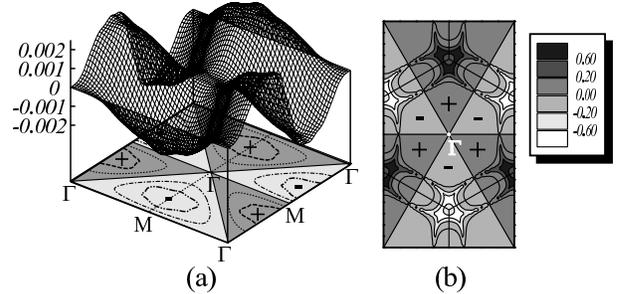}
\end{center}
\caption{(a) The gap function $\Del(k)$ with the $f$-wave triplet pairing 
($B_1$) obtained at $n=1.35$, $t'/t=0.23$, $U/t=1.8$ and $T/t=0.01$.
(b) The contour plot of $\F(k)=|\G(k)|^2\Del(k)$.
The absolute value is very large near K point on the Fermi surface.}
\label{fig:del}
\end{figure}
Fig.\ref{fig:del} shows $\Delta(k)$ and $\F(k)$
at $n=1.35$, $t'/t=0.23$, $U/t=1.8$ and $T/t=0.01$.
The symmetry of the gap function clearly possesses 
the $f$-wave ($B_1$) symmetry.
The line node crosses each Fermi pocket around M point 
in Fig.\ref{fig:fermi}(d).
The absolute value of $\F(k)$ becomes large near the Fermi surface 
around K point.
This is because the Fermi velocity becomes small near the van Hove points.
Thus, in this symmetry, the large superconducting gap is open 
in the vicinity of the van Hove points, and the energy gain is large.
Therefore, this solution is very stable.

In the case of $t'=0$, the eigen values are $\sim 0.05$ at most, 
and we can not expect superconductivity with high $\Tc$.
Rather, it means that the effect of the frustration in the ideal 
triangular lattice suppresses any instabilities.
The introduction of $t'$ is crucial to avoid such frustration.

\begin{figure}[t]
\begin{center}
\includegraphics[width=6cm]{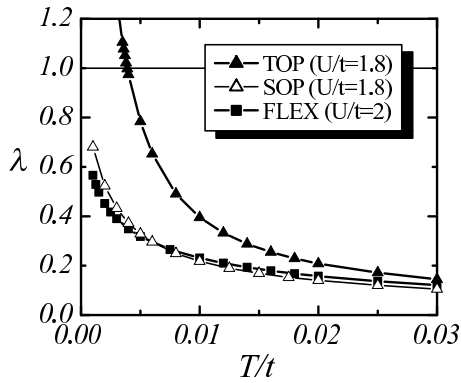}
\end{center}
\caption{The temperature dependence of eigen values for the $f$-wave pairing 
($B_1$) at $n=1.35$ and $t'/t=0.23$;
the thick line with black triangles for the third-order calculation,
the thin line with white triangles for the second-order calculation and
the thick line with black squares for the FLEX approximation.}
\label{fig:Tc}
\end{figure}
In Fig.\ref{fig:Tc}, we display $T$-dependence of eigen values 
for the $f$-wave triplet pairing at $n=1.35$, $t'/t=0.23$ and $U/t=1.8$.
For comparison, we have also evaluated the gap equation
in the second-order perturbation expansion (Eqs.(\ref{eq:sig2}) and 
(\ref{eq:del}a)), and in the FLEX approximation 
($U/t=2$).~\cite{rf:Kuroki,rf:Review}
In the both calculations, the same $f$-wave symmetry as discussed above is 
the most stable.
The eigen values are illustrated together in Fig.\ref{fig:Tc}.
It is interesting that the eigen value in the second-order calculation
is about the same as that in FLEX.
The pairing interaction in FLEX is given by the random phase approximation
of $\bar\chi_0(q)$ composed of the dressed Green's function.
In this case of $U/t=2$, the Storner-type enhancement factor 
$1/(1-U\bar\chi_0(q))$ is about 30.
Simply speaking, this provides the pairing interaction
30 times larger than that in the second-order calculation.
However, this enhancement also yields a strong depairing effect 
by the normal self-energy.
The strong depairing effect suppresses $\Tc$ too much.
On the other hand, the third-order calculation provides
finite $\Tc$ for the moderate $U/t=1.8$.
The third-order scattering process, especially including the particle-paritcle
ladder, weakens the depairing effect, which is over-estimated in FLEX.
Simaltaniously, it provides the strong attractive interaction 
for the spin-triplet pairing.
Thus, vertex correction terms other than simple one-boson propagation terms
are important for the spin-triplet superconductivity, 
like the case in Sr$_2$RuO$_4$.~\cite{rf:Nomura}
In this case, the convergence of solutions with respect to $U$ is very good.
Both the second- and third-order processes cooperatively stabilize 
this solution.
The transition temperature takes a reasonable value $\Tc \sim 0.005t \sim 5$K 
for $10t \sim 1$eV.~\cite{rf:Singh}
Thus, this $f$-wave triplet pairing ($B_1$) is a possible candidate
for the CoO$_2$ superconductor.
In fact, since $B_1$ pairing state is a non-degenerate one with line nodes, 
different from 2D representations $E_1$ and $E_2$, then we can naturally
explain the line-node like behavior observed in $1/T_1$.

In conclusion, we have investigated possibility of unconventional 
superconductivity in the repulsive Hubbard model on the 2D triangular lattice
with the next neighbor hopping $t'$.
We have evaluated the gap equation using the third-order perturbation 
expansion with respect to the on-site repulsion $U$.
We have found that the $f$-wave spin-triplet pairing state ($B_1$) is 
the most stable in the wide range of $t'$ and the electron density $n$ 
more than half filling.
In this case, the introduction of $t'$ is crucial to avoid the frustration
in the triangular lattice, and move the van Hove points 
near the Fermi surface around K point.
The bare spin susceptibility shows the broad peak around $\Gamma$ point.
These conditions stabilize the $f$-wave triplet pairing.
From comparison with the second-order calculation and the FLEX approximation,
the third-order vertex corrections are important 
for the spin-triplet superconductivity, 
rather than the remarkable ferromagnetic spin fluctuation.
This is also the case in Sr$_2$RuO$_4$.~\cite{rf:Nomura}
The $f$-wave triplet pairing ($B_1$) is the primary candidate
for the superconductivity in Na$_{0.35}$CoO$_2$.1.3H$_2$O,
as far as the van Hove points near the Fermi surface are located 
around K points.
We can expect the case from the recent LDA band 
calculation.~\cite{rf:Singh,rf:Kunes}
In addition, the large relaxation rate and uniform susceptibility
suggest high density of states 
at the Fermi level.~\cite{rf:Fujimoto,rf:Ishida}
These support our theoretical proposal.
At the present, in order to obtain a more reliable result about 
the possibility, we proceed with a further investigation 
using the band structure on the basis of the LDA band calculation.


The authors thank K. Ishida for useful information in the NQR experiments.
One of them (H.I.) also thanks S. Fujimoto for valuable discussions.
This study is financially supported by 
Grant-in-Aid for Scientific Research Areas (C)
from the Ministry of Education, Science, Sports, Culture, and Technology.



\begin{thebibliography}{99}
\bibitem{rf:calc}
We take 192 $\times$ 192 $k$-meshes for twice space of 
the first Brillouin zone and the cut-off $\sim 16t$ of Matsubara frequencies.
\bibitem{rf:gapeq}
This equation has been obtained by the perturbation expansion with respect to 
$\frac{U}{2}\sum_i n_i n_i=U\sum_i(n_{i\uparrow} n_{i\downarrow}+n_i/2)$
clearly isotropic in a spin space, 
instead of $U\sum_i n_{i\uparrow} n_{i\downarrow}$.
By considering the parity symmetry in $\F(k)$ for the spin-singlet and 
triplet pairs, usual equations are restored.
For instance, Eq.(\ref{eq:del}b) cancels out for the spin-triplet pairing.
This is because the Pauli principle is perfectly fulfilled
in the perturbation expansion.
\end{thebibliography}
\end{document}